\documentclass[twocolumn,aps,prl,superscriptaddress]{revtex4}
\usepackage{amssymb,bbold,dsfont}
\usepackage{amsmath}
\usepackage{color}
\usepackage{graphicx}
\usepackage{marvosym}
\usepackage{ulem}
\usepackage{notoccite}

\usepackage{comment}
\usepackage{pifont}

\usepackage{bbm}

\usepackage{graphicx,tabularx}
\usepackage{dcolumn}
\usepackage{bm}
\usepackage{xcolor}

\usepackage{marvosym}

\usepackage{titlesec}
\newcommand{\be}{\begin{equation}}
\newcommand{\ee}{\end{equation}}

\newcommand{\p}{\partial}
\newcommand{\la}{\langle}
\newcommand{\ra}{\rangle}

\renewcommand{\Re}{{\rm Re}\,}
\renewcommand{\vec}[1]{{\bf #1}}

\begin{document}

\title{Polariton-drag enabled quantum geometric photocurrents in high symmetry materials}
\author{Ying Xiong}
\affiliation{Division of Physics and Applied Physics, Nanyang Technological University, Singapore 637371}
\author{Li-kun Shi}
\affiliation{Max Planck Institute for the Physics of Complex Systems, 01187 Dresden, Germany}
\author{Justin C.W. Song}
\email{justinsong@ntu.edu.sg}
\affiliation{Division of Physics and Applied Physics, Nanyang Technological University, Singapore 637371}

\begin{abstract}
Lowered symmetry enables access to a wide set of responses not typically accessible in high symmetry materials. Prime examples are time-reversal forbidden quantum geometric photocurrent responses 
(e.g., linear injection and circular shift photocurrents) that are thought to vanish in non-magnetic materials. Here we argue that polariton-drag processes enable to unblock such quantum geometric photocurrents even in non-magnetic and centrosymmetric materials. Strikingly, we uncover how a cooperative effect between finite $\vec q$ irradiation and the Fermi surface position leads to a polariton selective photoexcitation (PSP). PSP enables to directly address carriers within tight momentum resolved windows of the Fermi surface to yield giant enhancements of quantum geometric photocurrents. This selectivity enables to directly track momentum resolved quantum geometric quantities along the Fermi surface providing a new tool to interrogate the quantum geometry of high symmetry materials.  

\end{abstract}

\maketitle

Quantum geometry can play an essential role in light-matter interaction. A prime example are bulk rectified currents such as the injection and shift photocurrents: these have strength determined by quantum geometric quantities (e.g., Berry curvature), and, as such, are now actively used as sensitive probes of the structure of Bloch wavefunctions in quantum materials~\cite{Morimoto2016, deJuan2017, Sodemann2015, Qiong2019, Watanabe, Nagaosa2020, Rappe, Ma_review}. Access to such photocurrents, however, requires lowered symmetry. For instance, circular shift (CS) and linear injection (LI) photocurrents are odd under time-reversal, $\mathcal{T}$, as well as inversion, $\mathcal{P}$. They are thought to only manifest in parity violating magnetic materials~\cite{Watanabe, Nagaosa2020,BingHaiYanNatComm, Qian2020} such as antiferromagnets. Consequently, the quantum geometric quantities associated with LI/CS photocurrents (e.g., quantum metric/circular shift vector) are typically inaccessible to photocurrent probes in high symmetry materials.

\begin{figure} [t!]
    \centering
    \includegraphics[scale=0.54]
{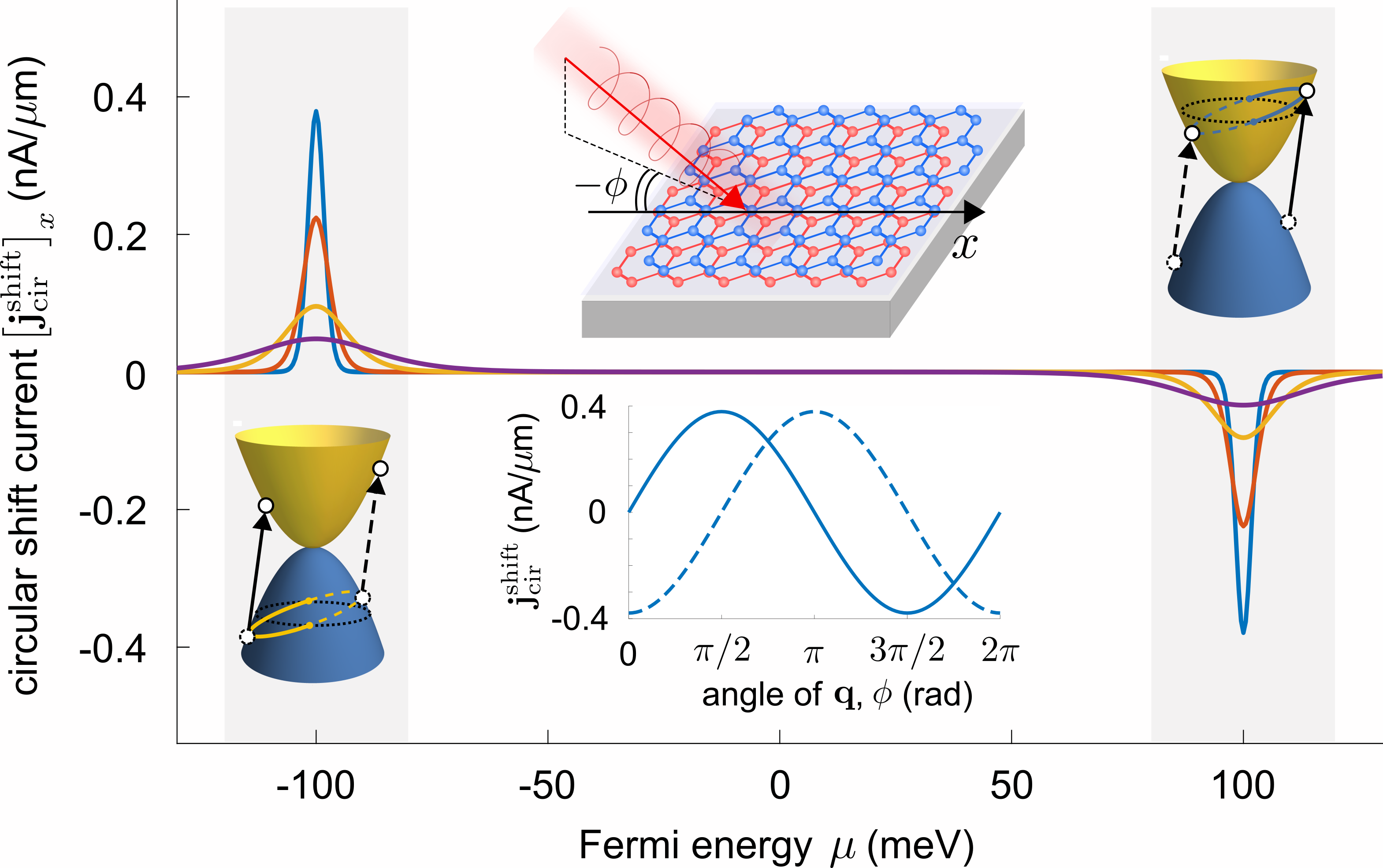}
    \caption{PD charge circular shift photocurrent in BLG displaying resonant like features close to $\mu = \pm \hbar \omega/2$; these arise from (inset) polariton-selective photoexcitation (PSP) wherein carries within a momentum window are excited, see also Fig. 2 and 3. Blue, orange, yellow and purple curves are obtained at temperatures 10 K, 20 K, 50 K and 100 K. Parameters used: $\hbar \omega = 200\, {\rm meV}$, $|\vec q| = 0.001 {\rm nm}^{-1}$ corresponding to that to free space photons, $|\vec E| = 0.05 \; {\rm V/\mu m}$, and $\phi = \pi/2$. (inset, middle upper panel) Schematic diagram of BLG irradiated by an oblique incidence of light. (inset, middle lower panel) Circular shift current as a function of $\phi$ for $\mu = -100$ meV. The solid (dashed) lines indicate $\left[ \vec j_{\rm cir}^{\rm shfit} \right]_x$ ($\left[ \vec j_{\rm cir}^{\rm shfit} \right]_y$).} 
    \label{fig1}
\end{figure}

Here we consider a different strategy to achieve lowered symmetry: by employing the spatial structure of electromagnetic (EM) fields (e.g., in nanophotonics~\cite{Kaminer2018, Koppens2015, Koppens2018} or under oblique incident irradiation). A case in point is the {\it drag} induced by photons or polaritons (e.g. propagating plasmon with a finite $\vec q$). In the polariton/photon-drag (PD) processes, the finite $\mathbf{q}$ momentum structure of travelling EM fields can induce non-vertical interband transitions. Indeed, exploiting PD has a long history: e.g., photon drag via direct optical transfer~\cite{Danishevskii, Gibson, Entin, Maysonnave, Shalygin, Ivchenko, Ganichev-Prettl, Glazov-review} or indirect processes~\cite{Ganichev-Prettl, Glazov-review, Plank} can be used to drive photocurrents, nanophotonic confinement can enable access to multipolar transitions~\cite{Rivera}. 

Here we argue that PD strategies can also be used to induce non-vertical interband transitions and bulk CS and LI photocurrents even in non-magnetic and inversion symmetric materials. Interestingly, PD strategies do not activate all charge quantum geometric photocurrents: we find PD linear shift and circular injection photocurrents still vanish in centrosymmetric and non-magnetic materials. This delineation highlights the central role quantum geometry plays in PD photocurrents: PD photocurrents depend on both quantum geometry and drag-induced velocity.

\begin{table*}
\centering
 \begin{tabularx}{\textwidth}{m{11em} >{\centering\arraybackslash}m{3.7cm} >{\centering\arraybackslash}m{3.7cm} >{\centering\arraybackslash}m{3.3cm} >{\centering\arraybackslash}m{3.3cm}  }
 \hline\hline\vspace{2mm}
 charge photocurrent  & linear injection & circular injection & linear shift & circular shift \\ [0.5ex] 
 \hline
 \vspace{2mm}
$\mathcal{P}$-symmetry & $\p_t \vec j_\theta(\vec q)= -\p_t \vec j_\theta(- \vec q)$ & $\p_t  \vec j_{\rm cir}  (\vec q) = - \p_t  \vec j_{\rm cir}  (- \vec q)$  & $\vec j_\theta(\vec q) = - \vec j_\theta(- \vec q) $ & $\vec j_{\rm cir} (\vec q) = - \vec j_{\rm cir} (- \vec q) $ \\
$\mathcal{T}$-symmetry & $\p_t \vec j_\theta(\vec q) = - \p_t \vec j_\theta(- \vec q)$ & $\p_t \vec j_{\rm cir}  (\vec q) =  \p_t \vec j_{\rm cir}  (- \vec q)$ & $\vec j_\theta(\vec q) =  \vec j_\theta (- \vec q)$ & $\vec j_{\rm cir} (\vec q) = - \vec j_{\rm cir} (- \vec q)$ \\\\
\hline
\vspace{2mm}
$\vec q \neq 0$ ($\mathcal{T}$ \& $\mathcal{P}$ symmetry)& \checkmark &\ding{53} & \ding{53} & \checkmark \\
$\vec q=0$ ($\mathcal{T}$ symmetry only) & \ding{53} & \checkmark & \checkmark & \ding{53} \\\\
 \hline \hline
\end{tabularx}
\label{tab:charge}
\caption{Symmetry relations for PD charge shift and injection photocurrents. Photocurrents for linear polarized irradiation are denoted $\theta$ whereas helicity dependent photocurrents are denoted ``cir''. We find that PD $\vec q \neq 0$ LI and CS charge photocurrents are allowed in both $\mathcal{T}$ and $\mathcal{P}$-preserving materials (indicated by ticks, third row). In contrast, when $\vec q=0$ LI and CS photocurrents vanish in $\mathcal{T}$-preserving but $\mathcal{P}$-breaking materials. }
\end{table*}

\begin{figure} [t!]
    \centering
    \includegraphics[scale=0.54]
{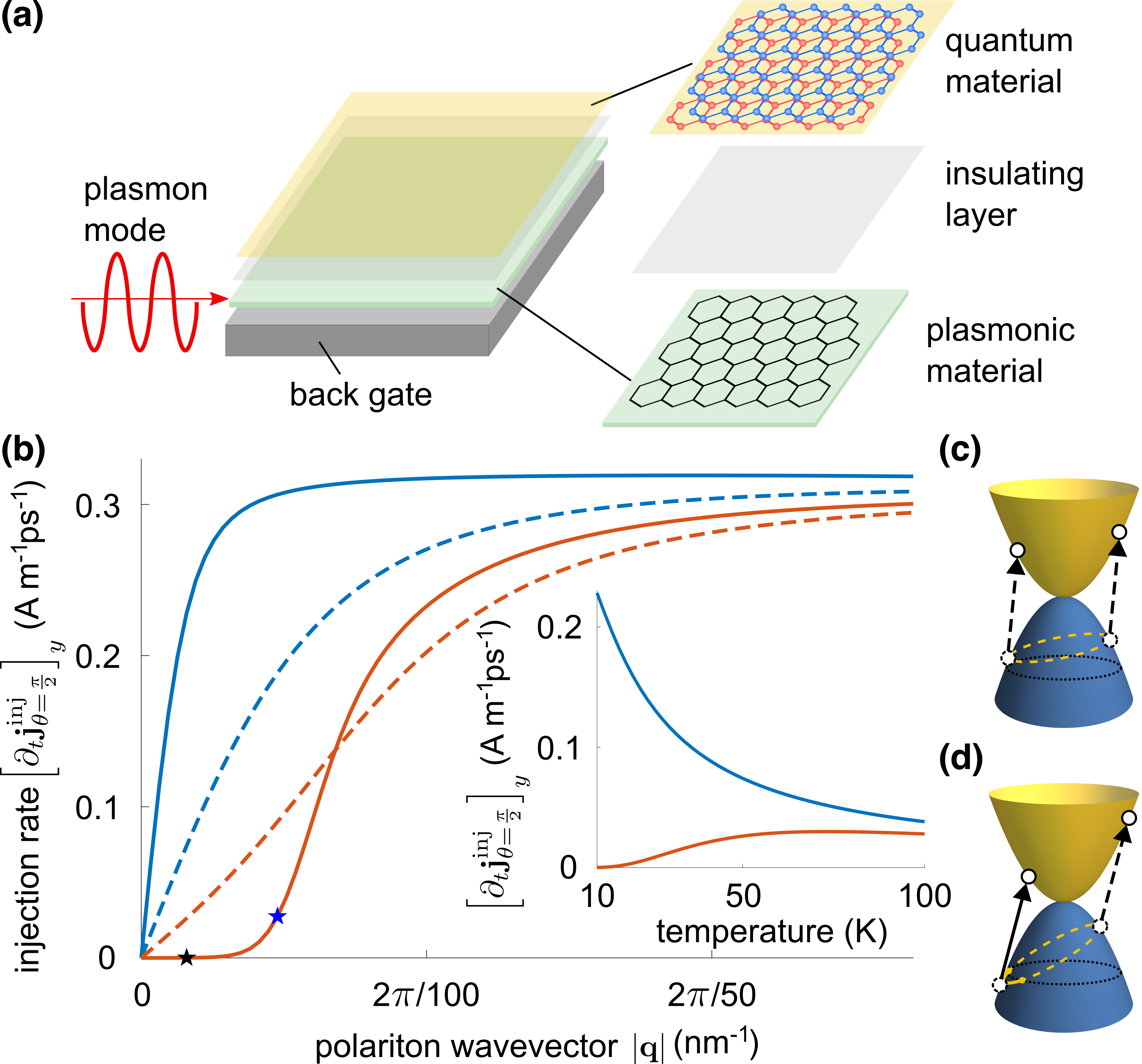}
    \caption{PD charge LI photocurrent in BLG. (a) Schematic diagram of a hybrid quantum material/host plasmonic material system, in which the electromagnetic field of the plasmons in the proximal host material (green) induces non-vertical interband transitions and PD photocurrents in the quantum material (yellow). (b) $|\vec q|$ dependence of PD LI photocurrent for Fermi energies $\mu = - 100$ meV (blue) and $\mu = -110$ meV (orange). 
    The solid and dashed lines denote temperatures of $10$ K and $50$ K. The black and blue stars correspond to $|\vec q| = 0.01\; {\rm nm^{-1}}$ (smaller than $ q_c$) and $|\vec q| = 0.03 \; {\rm nm^{-1}}$ (i.e. $|\vec q| \gtrsim q_c$). (inset) Temperature dependence of the PD LI photocurrent at $|\vec q| = 0.01 \;{\rm nm^{-1}}$ for different Fermi energies (colour code is the same as in main panel). We have set $|\vec E| = 0.05 \; {\rm V/\mu m}$ and $\phi = \pi/2$. (c, d) Schematic illustration of the transition contours (yellow) for $|\vec q|< q_c$ (c) and $|\vec q| \sim q_c$ (d). The yellow solid (dashed) lines indicate the occupied (unoccupied) section of the transition contour. The black dotted line indicates the Fermi surface. The injection photocurrent can be estimated from the injection rate by accounting for the relaxation or accumulation time, see description in text.}
    \label{fig2}
\end{figure}

Surprisingly, we find that PD can enable polariton selective photoexcitation (PSP) of carriers: i.e. by tuning both the polariton energy and its wavevector, only carriers within a selective window of momentum and energy are  photoexcited. As we explain below, PSP produces a rich phenomenology including resonant enhancements and Fermi surface dependent photocurrents that arise from interband transitions (see Figs.~\ref{fig1} and~\ref{fig2}). Importantly, when the momentum selective window of PSP is tightened, it can enable a photocurrent probe of momentum resolved quantum geometric quantities. 

A striking platform to realise strong PD photocurrents are hybrid plasmonic heterostructures~\cite{Koppens2011,Koppens2015,Koppens2018,Kaminer2018}, where a quantum material is placed on top of a plasmonic material (Fig.~\ref{fig2}a). In these, oblique incident light excites the plasmons in the plasmonic material, and the propagating EM field of the plasmon in turn induces a PD current in the quantum material. The wavevector of the plasmonic field can be tuned by dielectric constant of the substrate~\cite{Koppens2011} or by nanophotonic engineering~\cite{Koppens2015, Koppens2018}. Hybrid plasmonic heterostructures enable to achieve large $\vec q$-wavevectors far larger than that of free space, and, as we explain below, enhance PD photocurrents. 

\begin{figure*} [t!]
    \centering
    \includegraphics[scale=0.5]
{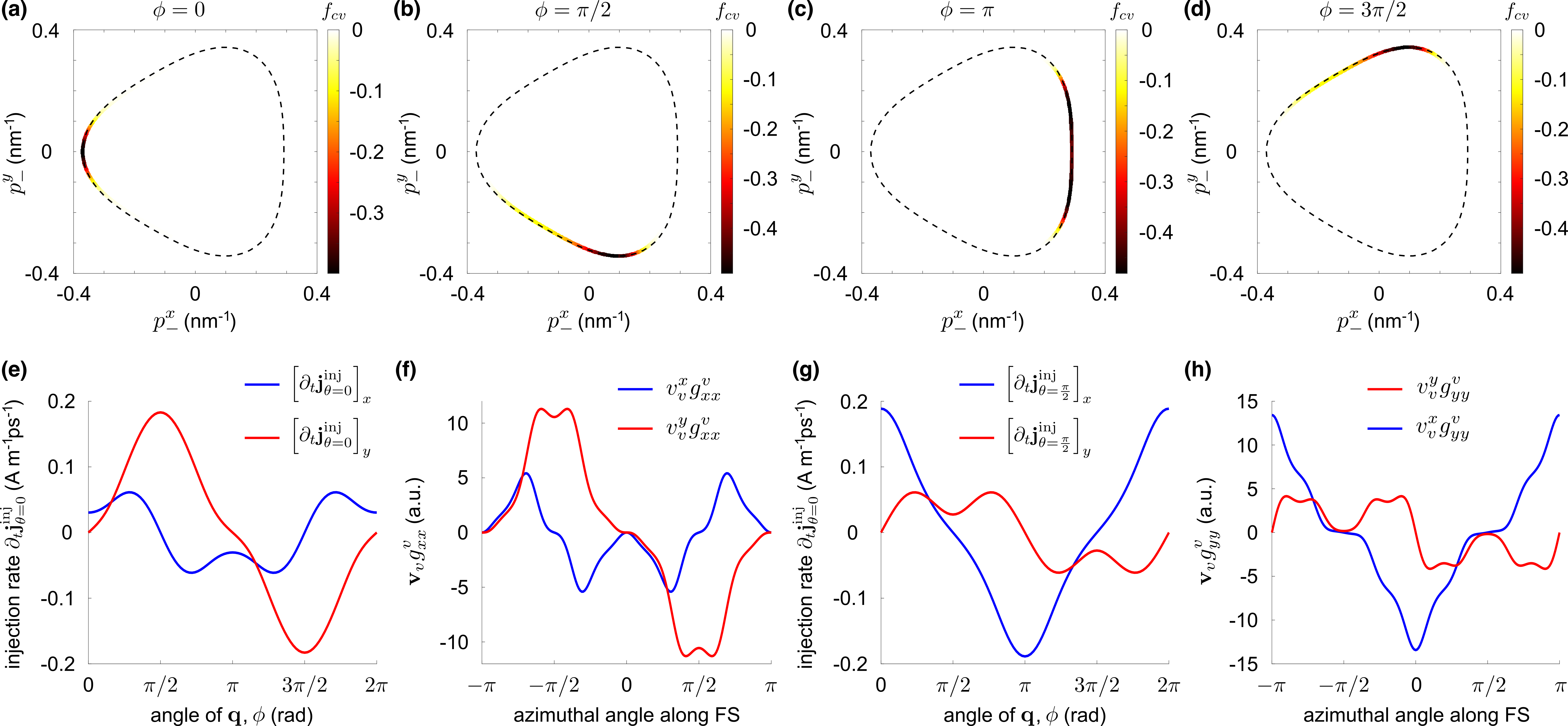}
    \caption{PD photocurrent as a momentum resolved tool to probe quantum geometry. (a-d) Partial excitation of charge carriers near the Fermi surface in k-space (coloured segment). Here $\vec p_- = \vec p - \vec q/2$. (e,g) LI photocurrents as a function of $\phi$ at a fixed magnitude $|\vec q| = 0.03\; {\rm nm^{-1}}$ and $\mu = -110$ meV for $x$-polarised (e) and $y$-polarised (g) light in BLG. Here we have used $T=$ 10 K and $|\vec E| = 0.05\; {\rm V/\mu m}$. The photocurrents enable to track the corresponding quantum metric dipoles shown in (f) and (h) along the Fermi surface (FS); here we have summed over both $K$ and $K'$ valleys in Eq.~(\ref{eq:H}). We note that for $\phi=0$, the charge carriers close to azimuthal angle $-\pi$ along the FS are sampled.} 
    \label{fig3}
\end{figure*}

As a concrete illustration, we show a PSP protocol for quantum geometric PD photocurrents in bilayer graphene (BLG) -- a centrosymmetric and non-magnetic material. We find PSP in BLG can induce large nonlinear susceptibilities with magnitudes comparable to that of ferroelectric materials~\cite{Qian} (where inversion symmetry broken) for values of $\vec q$ that can be readily achievable in hybrid plasmonic heterostructures. Strikingly, we find PD LI photocurrents tracks the momentum-resolved quantum metric dipole along the Fermi surface (see Fig.~\ref{fig3}). This demonstrates the power of polariton-drag processes in unblocking and amplifying quantum geometric photocurrents.

{\it PD injection and shift photocurrents.} 
We begin by considering a material irradiated by incident finite-$\vec q$ EM fields with electric field profile $\boldsymbol{\mathcal{E} }(\vec r, t) = (1/2) \sum_\pm \vec E_\pm e^{\pm i\vec q \cdot \vec r \mp i \omega t}$  with $\vec E_\pm$ the complex electric field amplitude where $\vec E_+ = (\vec E_-)^* = \vec E$. The oscillating EM fields induce real interband electronic transitions. As a concrete demonstration and for clarity and brevity of presentation, we focus on a two-band system where EM radiation induces transitions between the conduction $c$ and the valence $v$ bands. Considering momentum and energy conservation, EM radiation induces non-vertical transitions between pairs of Bloch states $|u_v (\vec k_-)\ra$ and $|u_c( \vec k_+)\ra$, where $\vec k_- = \vec k - \vec q/2$ and $\vec k_+ = \vec k + \vec q/2$. The transition rates can be readily calculated by Fermi's golden rule: $W_{i\to f}^\pm = (2 \pi/\hbar) |V_{i \to f}^\pm |^2 f_i (1-f_f) \delta(\epsilon_f - \epsilon_i \mp \hbar \omega )$, where $V_{i\to f}^\pm = e/ (2 \omega) \la f |\vec E^\pm \cdot \hat{ \boldsymbol{\nu}} | i \ra$ is the interband transition matrix element, $\hat{ \boldsymbol{\nu}}$ is the velocity operator, and $f_{i(f)}$ and $\epsilon_{i(f)}$ are the electron distribution function and energy for the initial (final) states respectively. Accounting for the changes to electron position and velocity upon interband transition directly produce (interband) quantum geometric photocurrents~\cite{Nagaosa2020}. In the following, we focus on the PD photocurrents induced by interband transitions. These are expected to dominate in the high frequency regime when the polariton frequency $\omega$ is much larger than the carrier scattering rate~\cite{Silkin}.

To see this, we first examine the shift current that arises from the real space displacement of electrons (charge $e<0$) that undergo interband transitions~\cite{Sipe2000, vonBaltz1981, Nagaosa2020}. Accounting for the transition rate, the finite-$\vec q$ shift current is~\cite{Likun} 
\be\label{eq:shift}
\vec j^{\rm shift} (\vec q) = C \sum_{\vec k} \rho(\vec k, \vec q) |  \vec E \cdot \boldsymbol \nu_{cv}(\vec k, \vec q)|^2 \vec R (\vec k, \vec q), 
\ee
where $C = -e^3 \pi/(2 \hbar \omega^2)$, the occupation factor $\rho (\vec k, \vec q) = f_{cv} (\vec k, \vec q) \delta (\epsilon_{cv} (\vec k, \vec q) - \hbar \omega)$, the Fermi function difference is $f_{cv} (\vec k, \vec q) = f(\epsilon_{c} (\vec k_+)) - f(\epsilon_{v}(\vec k_-))$ with the interband transition energy $\epsilon_{cv} (\vec k, \vec q) = \epsilon_c (\vec k_+) - \epsilon_v (\vec k_-)$, the velocity matrix element is $\boldsymbol \nu_{cv} (\vec k, \vec q) = \la u_c (\vec k_+) | \hat{\boldsymbol{\nu}} (\vec k) | u_v (\vec k_-) \ra$, and $\vec R (\vec k, \vec q)$ is the real-space displacement~\cite{Morimoto2016, LikunPRB, Nagaosa2020}, also called the shift vector, when a valence electron transits to the conduction band. It is directly related to a Pancharatnam-Berry (geometric) phase $\vec R (\vec k, \vec q) = \lim_{\delta \vec k \to 0} \nabla_{\delta \vec k} \arg \mathcal W (\vec k, \delta \vec k, \vec q)$ accrued during the transition (see Supplemental Material, {\bf SM}~\cite{SM}):  
\begin{align}
\mathcal W (\vec k, \delta \vec k, \vec q) & = \la u_v (\vec k_-) | u_v(\vec k'_-) \ra [ \hat{\vec e} \cdot \la u_v (\vec k'_- )| \hat{\boldsymbol \nu} | u_c (\vec k'_+) \ra ]  \nonumber \\ & \cdot \la u_c (\vec k'_+) | u_c (\vec k_+) \ra \la u_c (\vec k_+ ) | u_v (\vec k_-) \ra, 
\label{eq:shiftvector}
\end{align}
where $\hat{\vec e}$ is the polarization, $\vec k'_- = \vec k_- + \delta \vec k$, and $\vec k'_+ = \vec k_+ + \delta \vec k$. 

In the same fashion, the injection current rate arises from a change of velocity when a carrier undergoes interband transitions~\cite{Nagaosa2020} and can be written as 
\be\label{eq:inj}
 \p_t \vec j^{\rm inj} (\vec q) =  C\sum_{\vec k} \rho(\vec k, \vec q) | \vec E \cdot \boldsymbol \nu_{cv}(\vec k, \vec q)|^2 \boldsymbol \Delta (\vec k, \vec q), 
\ee
where $\boldsymbol{\Delta} (\vec k, \vec q) =   \vec v_{c} (\vec k_+) - \vec v_{v} (\vec k_-) $ is the change in carrier velocity. In the same fashion as above, the transition matrix elements are closely related to an interband quantum geometric tensor. As we will see below, this fact, together with PSP, will enable to probe the momentum resolved quantum geometry of Bloch bands.

We note that Eq.~(\ref{eq:inj}) describes a rate of change of injection current. In physical situations, this accumulation of current is often cut by a finite relaxation time or in the case of ultra-short pulses of EM radiation where the pulsewidth duration is shorter than relaxation time, by the pulsewidth. As such, the injection current can be estimated as $\vec j^{\rm inj} (\vec q) = \tau \p_t \vec j^{\rm inj} (\vec q)$~\cite{Rees,Holder,deJuan2017, grushin}, where $\tau$ is an effective time over which the injection photocurrent relaxes/accumulates. In steady-state measurements, $\tau$ is often approximated by the momentum relaxation time of the photoexcited hot-carriers~\cite{deJuan2017, Rees}; we note, parenthetically, that understanding the precise interplay between relaxation and quantum geometric photocurrents is a subject of current intense research~\cite{Matsyshyn, Holder}. The relaxation time can even be band and $\vec k$ dependent~\cite{Danishevskii, Ivchenko, Ganichev-Prettl}. In what follows, to highlight the PSP effect, we will focus on the ultrafast photocurrent regime.

{\it Unblocking time-reversal forbidden photocurrents.} As we now argue, both $\vec j^{\rm shift} (\vec q)$ and $\vec j^{\rm inj} (\vec q)$ in Eq.~(\ref{eq:shift}) and Eq.~(\ref{eq:inj}) possess markedly different symmetry properties as compared to their vertical transition counterparts. We perform a symmetry analysis
to obtain the PD photocurrent symmetry properties shown in Table 1, see {\bf SM} for details. In populating the table, we have denoted photocurrents $\vec j_\theta$ arising from linearly polarized light $\vec E = \vec E^\theta = E (\hat{\vec x} \cos \theta + \hat{\vec y} \sin \theta) $ with the subscript index $\theta$. In analysing the circularly polarised irradiation, we have focused on the photocurrent $\vec j_{\rm cir}$ that depends on light helicity $\eta$ [with electric field $\vec E^\eta = E (\hat{\vec x} + i \eta \hat{\vec y})$].

Of particular note are the LI and CS photocurrents. While forbidden when $\vec q=0$ in $\mathcal{T}$ invariant non-magnetic materials, non-vertical transitions (PD activated) when $\vec q \neq 0$ enable to generate finite PD LI and CS photocurrents even in materials with both $\mathcal{T}$ and $\mathcal{P}$ symmetries (third row). This is because LI/CS photocurrents display an odd parity as $\vec q \to -\vec q$ for either $\mathcal{T}$ and $\mathcal{P}$ symmetries: $\vec q$ controls the direction of the PD LI/CS photocurrent generated.

Interestingly, finite $\vec q$ circular injection and linear shift charge photocurrents {\it vanish} in materials possessing both $\mathcal{T}$ and $\mathcal{P}$ symmetries: not all photocurrents are enabled by finite $\vec q$; this mirrors a similar vanishing in $\mathcal{PT}$ symmetric parity-violating magnets at $\vec q=0$~\cite{Watanabe, Nagaosa2020}. We note that while here we have concentrated on charge photocurrent response, PD {\it spin} photocurrents are expected to have different transformation properties from that of Table 1~\cite{Likun}. Lastly, we note that while we have focused on interband photocurrents, finite $\vec q$ may also unblock intraband photocurrents that can depend on extrinsic scattering processes in non-magnetic and centrosymmetric metals~\cite{Silkin}.

{\it PD CS and LI photocurrents in BLG.} As a concrete demonstration of how non-vertical transitions unblock quantum geometric photocurrents, we examine CS and LI photocurrents in gapless BLG. Notably, BLG is a centrosymmetric semimetal that preserves $\mathcal T$-symmetry; its low energy Hamiltonian can be written as $H(\vec p) = H_0 (\vec p) + H_w (\vec p) $ \cite{Koshino}, where 
\begin{align}\label{eq:H}
&H_0 (\vec p) = -\frac{\hbar^2 }{2m} \left[ (p_x^2 - p_y^2)  \sigma_x + 2 \zeta p_x p_y \sigma_y \right], \nonumber \\
& H_w (\vec p) = \hbar v_3 (\zeta p_x \sigma_x - p_y \sigma_y).
\end{align}
Here $\vec p = \vec k - \vec K_\zeta$ is the Bloch wavevector measured from $K_\zeta$ points, $\zeta = \pm$ is the valley index, and $m$ is the effective mass. $H_w (\vec p)$ describes trigonal warping, consistent with BLG's three-fold rotational symmetry $C_3^z$. Additionally, BLG also possesses mirror axes (e.g., $y$-axis act as a mirror plane).

We first examine the PD CS photocurrent. We evaluate Eq.~(\ref{eq:shift}) for a circularly polarized beam ($\hbar \omega = 200 $ meV)
with in-plane photon wavevector $\vec q \parallel \hat{\vec y}$ along a mirror axis in BLG. This yields a sizeable PD $\vec j_{\rm cir}^{\rm shift}$ in Fig.~\ref{fig1}. We note that point group symmetries can greatly constrain the direction of the PD photocurrents for a given polarisation. To see this, consider mirror symmetry $\mathcal M_y: (x,y) \to (-x, y)$ with the polariton wavevector $\vec q$ along the mirror axis. For circularly polarised light, we find that the momentum resolved transition rate obeys $|\vec E^{\eta} \cdot \vec v_{cv}(\vec k, \vec q)|^2 = |\vec E^{-\eta} \cdot \vec v_{cv}(\mathcal M_y \vec k, \vec q)|^2$, while the shift vector satisfies (see {\bf SM} for detailed analysis) 
\be
R^\eta_x (\vec k, \vec q) = - R^{-\eta}_x (\mathcal M_y \vec k, \vec q), \quad R^\eta_y (\vec k, \vec q) = R^{-\eta}_y (\mathcal M_y \vec k, \vec q).
\ee
As a result, when $\vec q$ is directed parallel to mirror plane, we find the PD circular shift photocurrent $\vec j_{\rm cir}^{\rm shift}$ is transverse. This is verified in the numerical simulation for BLG, as shown in Fig.~\ref{fig1} inset.

Strikingly, PD CS photocurrents display large peaks centered at $\mu = \pm \hbar \omega/2$ (Fig.~\ref{fig1}). These resonant peaks arise from PSP: when the (tilted) interband transition energy contours [defined by $\delta (\epsilon_{cv}^\zeta (\vec p, \vec q) - \hbar \omega)$] intersect with the Fermi surface. In this, the combined action of the finite-$\hbar \vec q$ momentum transfer as well as the position of the Fermi surface ensures that only carriers in parts of the interband transition energy contours $\delta (\epsilon_{cv}^\zeta (\vec p, \vec q) - \hbar \omega)$ are excited [as captured by the joint occupation factor $\rho (\vec p, \vec q)$ in Eq.~(\ref{eq:shift})]. PSP induces a large asymmetry in sampling the circular shift vector (see {\bf SM}) to produce a giant enhancement of CS photocurrent.

Interestingly, the part of the interband transition energy contour that is excited depends directly on $\mu$:  
when $\mu$ is tuned from $-\hbar \omega/2 \to \hbar \omega/ 2$ the allowed excitations flip (see inset Fig.~\ref{fig1} and {\bf SM}) thereby sampling a different window of circular shift vector $\vec R^{\eta,\zeta} (\vec p, \vec q)$, where $\eta$ denotes the shift vector induced by light with helicity $\eta$. Indeed, this sampling is angle sensitive: by rotating azimuthal angle $\phi$, CS photocurrent similarly rotates [Fig.~\ref{fig1} (inset)] displaying a photocurrent that is locked to the symmetry breaking axis determined by $\vec q$. When $\mu$ is tuned away from $\pm \hbar \omega/2$, PD $\vec j_{\rm cir}^{\rm shift}$ falls steeply (Fig.~\ref{fig1}a); in this regime $\vec j_{\rm cir}^{\rm shift}$ vanishes due to the presence of $c, \; v$ band symmetry in Eq.~(\ref{eq:H}), see {\bf SM}. We note that such $c, \; v$ band symmetry is strongly broken by tuning the Fermi surface so that it intersects with the interband transition contour, leading to PSP and large photoresponse. 

PSP-induced peak features are ubiquitous for PD photocurrents and also extend to PD LI photocurrents. Indeed, similar peaks close to $\mu = \pm \hbar \omega/2$ have been predicted for oblique incident far-field linearly polarized light at low-temperature in graphene~\cite{Entin}. We note, in a similar fashion described above for PD CS photocurrents, the direction of PD LI photocurrent also exhibits a strong dependence of high-symmetry axes in the material; in particular, it is highly sensitive to how light polarisation is aligned with the mirror axes (see {\bf SM} for details). To see this, consider the case when $\vec q$ is along the mirror axis, so that  
\be
\Delta_x (\vec k, \vec q) = - \Delta_x (\mathcal M_y \vec k, \vec q), \quad \Delta_y (\vec k, \vec q) = \Delta_y (\mathcal M_y \vec k, \vec q).
\ee
For the special case of light polarised either parallel or perpendicular to the mirror axis, we have $|\vec E^{\theta} \cdot \vec v_{cv}(\vec k, \vec q)|^2 = |\vec E^{\theta} \cdot \vec v_{cv}(\mathcal M_y \vec k, \vec q)|^2$, yielding PD LI photocurrent flowing along $\vec q$. For light polarized away from these directions, mirror symmetry is broken and PD LI photocurrents need not flow purely along $\vec q$, see {\bf SM}.

In the following, we concentrate on a different regime for PD LI photocurrents: $\mu$ is detuned away from $\pm \hbar \omega/2$. In this detuned situation (red curve Fig.~\ref{fig2}b), small values of $|\vec q| \ll q_c$ do not produce an LI photocurrent since the transition contour does not intersect the Fermi surface (Fig.~\ref{fig2}c); here $q_c$ is a threshold wavevector at which the transition contour (defined by $\epsilon_{cv} (\vec k, \vec q) = \hbar \omega$) just intersects the Fermi surface (Fig.~\ref{fig2}d). In the conduction/valence band, the transition contour is given by $\epsilon_{c,v} (\vec k \pm \vec q/2)$. For small detuning, $q_c$ can be estimated as $q_c \approx 2| |\mu| - \hbar \omega/2|/\hbar \tilde v$ where $\tilde v = {\rm max} \left[ \tilde{\vec v}_{c,v} \cdot \hat{\vec q} \right]$, and $\tilde{\vec v}_{c,v} = (1/\hbar)  \left[ \nabla_{\vec k} \epsilon_{c,v} (\vec k) \big|_{\epsilon_{c,v} = \pm \hbar \omega/2} \right]$. When $|\vec q| \gtrsim q_c$, LI photocurrent rapidly turns on: this arises from a tight PSP window of photoexcited carriers. As illustrated in Fig.~\ref{fig2}b, for a detuning of $||\mu| -\hbar \omega/2| = 10$ meV, the LI photocurrent turns on at $q_c \approx 0.03 \; {\rm nm^{-1}}$ (blue star), which is about 30 times larger than that of free space light and can be readily achieved in graphene based plasmonic materials~\cite{Koppens2011}.

This behavior contrasts with that of $\mu=-\hbar \omega/2$ case (blue curve) where PD LI photocurrents flow even for arbitrarily small but finite values of $\vec q$ since $q_c = 0$: even small $\vec q$ produce a wide window (in momentum space) of PSP carriers. This difference in PSP windows for $|\mu|=\hbar \omega/2$ vs $\mu \neq \hbar \omega/2$ leads to contrasting temperature dependence. When $|\mu|=\hbar \omega /2$ PD LI photocurrents increase as temperature decreases. In contrast, when $\mu \neq \hbar \omega/2$, PD LI photocurrents display a complex $\vec q$-dependent temperature dependence since $\vec q$ controls the regions of the interband transition contour that dip below the Fermi surface. When $\vec q$ is further increased beyond $q_c$, the LI photocurrent saturates and become relative insensitive to temperature, see Fig.~\ref{fig2}b (solid vs dashed). Note that in Fig.~\ref{fig2}b, we have plotted the LI photocurrent for a range of wavevectors up to $|\vec q| = 0.17 \; {\rm nm^{-1}}$, which corresponds to a plasmonic field confinement of 170 times and can be achieved via nanophotonic engineering~\cite{Koppens2015, Koppens2018}.

Strikingly, when $\mu$ is detuned away from $|\mu | = \hbar \omega/2$ and by selecting $|\vec q| \approx q_c$ just at threshold, a highly momentum selective PSP window can be engineered (Fig.~\ref{fig3}a-d where the amplitude of $\rho(\vec p, \vec q)$ is plotted). At these $|\vec q|$ values, the transition contour just intersects the Fermi surface. As a result, PSP enables angle-selective (controlled by the direction of the polariton wavevector, $\phi$) excitation of carriers close to the Fermi surface (dashed black line); this mirrors means of momentum resolution found in angle-resolved photoemission. We note that when $\mu$ is tuned from the valence band to the conduction band, charge carriers from the opposite side of the Fermi surface are sampled. 

Here we have chosen a Fermi energy detuning of 10 meV away from $- \hbar \omega /2$ and the plasmon polariton wavevector $|\vec q| = 0.03 \; {\rm nm^{-1}}$. In this regime, $k_B T $ (employed in Fig.~\ref{fig3}) is much smaller than the detuning, allowing for a good momentum resolution of PSP. We note that in principle, such selective photoexcitation can also be achieved using wavevectors that are smaller (e.g., using free space photons). However, the corresponding detuning to achieve tight momentum resolution will be similarly smaller, making such selectivity highly sensitive to thermal broadening and easily smeared.

The tight window of PSP-induced excitation enables to probe momentum resolved quantum geometry near the Fermi surface. To see this, consider the PD {\it linear} injection current in Eq.~(\ref{eq:inj}) written out in component form as $\p_t \vec j^{\rm inj}_\theta (\vec q) = -e^3 \pi/(2 \hbar) \sum_{\vec k, a, b} \rho(\vec k, \vec q) \boldsymbol{\Delta} (\vec k, \vec q) G_{ba}^{cv} (\vec k, \vec q) E_b E_a^* $, where 
\be\label{eq:G}
G_{ba}^{cv} (\vec k, \vec q) = \Re \big\{ r_{cv}^{b} (\vec k, \vec q) [r_{cv}^{a} (\vec k, \vec q)]^* \big\}
\ee
is a generalised $\vec q$-dependent $c,v$ band resolved interband quantum metric (see {\bf SM}) with $r_{cv}^a (\vec k, \vec q)=\hbar\nu_{cv}^a (\vec k, \vec q)/ i [\epsilon_c(\vec k_+) - \epsilon_v(\vec k_-) ]$. Interestingly, in the two-band limit that we concentrate on [Eq.~(\ref{eq:H})] and when polariton wavevector is relatively small, $|\vec q| \ll |\vec k|$, $\boldsymbol{\Delta} (\vec k, \vec q) G_{ba}^{cv} (\vec k, \vec q)$ approximates the direct quantum metric dipole $\boldsymbol{\Delta} (\vec k,0) \Re \big[r_{cv}^b (\vec k,0) r_{vc}^a (\vec k,0) \big] = - 2 \vec v_v(\vec k) g_{ba}^v (\vec k)$~\cite{Nagaosa2020}, where in the last equality we specialized to BLG in Eq.~(\ref{eq:H}) and $g_{ba}^v(\vec k) dk_adk_b= 1- | \langle u_v (\vec k) | u_v(\vec k + d\vec k) \rangle|^2$ is the valence band quantum metric. For comparison with the numerical photocurrent simulation in Fig.~\ref{fig3}, here we explicitly write down $g_{xx}^v (\vec p)$ and $g_{yy}^v (\vec p)$ at each valley: 
\begin{align}
&g_{xx}^v (\vec p) = \frac{ \left[ \left( - \frac{\hbar^2}{m} p_x +\hbar v_3 \zeta \right) \sin \varphi_{\zeta, \vec p} + \frac{\hbar^2}{m} \zeta p_y \cos \varphi_{\zeta, \vec q} \right]^2}{4 [d(\vec p)]^2 }, \nonumber \\
&g_{yy}^v (\vec p) = \frac{ \left[ \frac{\hbar^2}{m} p_y \sin \varphi_{\zeta, \vec p} + \left( \frac{\hbar^2}{m} \zeta p_x + \hbar v_3 \right) \cos \varphi_{\zeta, \vec p} \right]^2}{4 [d(\vec p)]^2 }, 
\end{align}
where $d(\vec p) = \sqrt{d_1^2 + d_2^2} $ is the conduction band energy, $d_1 = -\frac{\hbar^2}{2m} (p_x^2 - p_y^2)+ \hbar v_3 \zeta p_x $, $d_2 = - \frac{\hbar^2}{m} \zeta p_x p_y - \hbar v_3 p_y$, and $\tan^{-1} \varphi_{\zeta, \vec p} = d_2 / d_1$. As we discuss below, the corresponding momentum resolved quantum metric dipoles (Fig.~\ref{fig3}f and h) provide a good estimate for the PD LI photocurrents for the two-band Hamiltonian in Eq.~(\ref{eq:H}). Of course, in a more general multiband Hamiltonian, PD LI photocurrents track the generalized interband quantum metric, see Eq.~(\ref{eq:G}).

Given the tight momentum selective window accessed in Fig. 3a-d, by fixing the magnitude of $\vec q$ while varying its direction, the LI photocurrents enables to track the quantum metric dipole $g_{ba}^v (\vec k) \vec v_v (\vec k)$ distribution along the Fermi surface. Indeed, Fig.~\ref{fig3}e-h provides a comparison between the LI photocurrents induced by $x$- and $y$-polarised light (Fig.~\ref{fig3}e and g respectively) and the corresponding quantum metric dipoles (Fig.~\ref{fig3}f and h) along the Fermi surface. We observe that the LI photocurrents capture the main features of the quantum metric dipole as a function of azimuthal angle along the Fermi surface. 
 
PD can be used as a ``knob'' to turn-on, control, and amplify quantum geometric photocurrents in a wide range of high symmetry materials even when either $\mathcal{P}$ or $\mathcal{T}$ or both symmetries are intact. Even as we have focussed on how PSP enables to probe momentum resolved quantum geometry, from an applied perspective, the selective excitation of carriers enables a novel means of amplifying non-linear susceptibilities: by exploiting PSP to selectively excite carriers with similar group velocities. As an example, we find PSP enhanced LI susceptibilities as high as $\eta_{yyy} (q) \sim 10^{10} - 10^{11} \, {\rm A}\, {\rm nm} {\rm V}^{-2} {\rm s}^{-1}$ in BLG (a $\mathcal{P}$ and $\mathcal{T}$ preserving material) comparable with those found in 2D ferroelectrics~\cite{Qian}.

\vspace{2mm}
\begin{acknowledgments}
{\bf Acknowledgements -- } We gratefully acknowledge useful conversations with Mark Rudner, Qiong Ma, Cheng Liang, Elbert Chia and Arpit Arora. This work was supported by Singapore MOE Academic Research Fund Tier 3 Grant MOE2018-T3-1-002 and a Nanyang Technological University start-up grant (NTU-SUG). 
\end{acknowledgments}

\clearpage

\newpage

\setcounter{equation}{0}
\setcounter{figure}{0}
\renewcommand{\theequation}{S\arabic{equation}}
\renewcommand{\thefigure}{S\arabic{figure}}

\renewcommand{\bibnumfmt}[1]{(#1)}

\setcounter{secnumdepth}{4}

\onecolumngrid

\begin{center}
\textbf{\large Supplemental Materials for ``Polariton-drag enabled quantum geometric photocurrents in high symmetry materials"}

\bigskip

Ying Xiong,$^1$ Li-kun Shi,$^2$ and Justin C.W. Song$^{1,*}$

\bigskip
$^1${\it Division of Physics and Applied Physics, Nanyang Technological University, Singapore 637371}

$^2${\it Max Planck Institute for the Physics of Complex Systems, 01187 Dresden, Germany}
\end{center}

\section{Geometric representation of shift and injection current}

\subsection{Geometric representation of shift current}

The shift photocurrent arises from a real-space displacement $\vec R (\vec k, \vec q)$ when an electron undergoes interband transition, see Eq.~(\ref{eq:shift}) in the main text. The finite-$\vec q$ shift vector $\vec R(\vec k, \vec q)$ can be written as~\cite{Likun}
\be\label{eq:shiftvector_old}
\vec R (\vec k, \vec q) = \vec A_c (\vec k_+) - \vec A_v (\vec k_-) - \nabla_{\vec k} \arg \left[\hat{\vec e} \cdot \boldsymbol \nu_{cv}(\vec k, \vec q) \right],
\ee
where $\vec A_n(\vec k) = \la u_n (\vec k) | i \nabla_{\vec k} | u_n (\vec k)\ra$ is the Berry connection and $\hat{\vec e}$ is the unit vector for the electric field polarisation. The shift vector is determined by the quantum geometry of the Bloch bands and the light polarisation. To see this, we rewrite $\vec R(\vec k, \vec q)$ as a derivative of the Pancharatnam-Berry phase obtained from the Wilson loop associated with the interband transition, as defined in Eq.~(\ref{eq:shiftvector}) of the main text. To uncover the geometric meaning of the shift vector, we first note that the Berry connection can be written: 
\be
 A_n^b (\vec k) = - \lim_{\delta k_b \to 0} \partial_{k_b} \arg \la u_n (\vec k) | u_n (\vec k+ \delta k \hat{\vec b}) \ra = \lim_{\delta k_b \to 0} \partial_{k_b} \arg \la u_n (\vec k + \delta k \hat{\vec b}) | u_n (\vec k) \ra, 
\ee
where $\hat{\vec b}$ is the unit vector in direction $b= \{ x, y\}$. 
On the other hand, the last term in Eq.~\ref{eq:shiftvector_old} can be rewritten as 
\be
- \nabla_{\vec k} \arg \left[ \hat{\vec e} \cdot \boldsymbol \nu_{cv}(\vec k, \vec q) \right] = \nabla_{\vec k} \arg \la u_v (\vec k_-) | \hat{\vec e} \cdot \hat{\boldsymbol \nu} | u_c (\vec k_+)\ra  = \lim_{\delta \vec k \to 0} \nabla_{\delta \vec k} \arg \la u_v (\vec k_- + \delta \vec k) |  \hat{\vec e} \cdot \hat{\boldsymbol \nu} | u_c (\vec k_+ + \delta \vec k)\ra
\ee
Therefore, the shift vector can be expressed as the gradient of the Panchanratnam-Berry phase of the Wilson loop 
\be\label{eq:r_Wilson}
\vec R (\vec k, \vec q) = \lim_{\delta \vec k \to 0} \nabla_{\delta \vec k}  \arg \mathcal W (\vec k, \delta \vec k, \vec q),
\ee
where 
\be\label{eq:Wilson}
\mathcal W (\vec k, \delta \vec k, \vec q) = \la u_v (\vec k_-) | u_v(\vec k'_-) \ra [ \hat{\vec e} \cdot \la u_v (\vec k'_- )| \hat{\boldsymbol \nu} | u_c (\vec k'_+) \ra ]  \la u_c (\vec k'_+) | u_c (\vec k_+) \ra \la u_c (\vec k_+ ) | u_v (\vec k_-) \ra
\ee
with $\vec k'_- = \vec k_- + \delta \vec k$, and $\vec k'_+ = \vec k_+ + \delta \vec k$. Here we have introduced $\la u_c (\vec k_+ ) | u_v (\vec k_-) \ra$ to complete the Wilson loop; we note that $\nabla_{\delta \vec k}  \arg \left[ \la u_c (\vec k_+ ) | u_v (\vec k_-) \ra\right] = 0 $ does not contribute to the shift vector. Eq.~(\ref{eq:r_Wilson}) and~(\ref{eq:Wilson}) provide a geometric interpretation of the shift vector, which corresponds to the gradient of the Pancharatnam-Berry associated with the interband transitions.

\subsection{Geometric representation of the injection current} 
In this section, we show that the injection photocurrents depend on the quantum geometric tensor of the material. To see this, we note that the injection current for an arbitrarily polarised light can be written as 
\begin{align}\label{eq:inj_qm}
\p_t \vec j^{\rm inj} (\vec q)&= C \sum_{\vec k, a, b} \rho(\vec k, \vec q) \boldsymbol \Delta (\vec k, \vec q) \nu_{cv}^b (\vec k, \vec q) [ \nu_{cv}^a (\vec k, \vec q) ]^* E_b E_a^* \nonumber \\
&= - \frac{e^3\pi }{2 \hbar} \sum_{\vec k, a, b} \rho(\vec k, \vec q) \boldsymbol \Delta (\vec k, \vec q) Q_{ba}^{cv} (\vec k, \vec q)  E_b E_a^*
\end{align}
Here we have defined $r_{cv}^a (\vec k, \vec q) = \nu_{cv}^a (\vec k, \vec q) / [i \omega_{cv} (\vec k, \vec q)] $ and $\omega_{cv} (\vec k, \vec q) = [\epsilon_c(\vec k_+) - \epsilon_v(\vec k_-) ] /\hbar$. In the second line of Eq.~\ref{eq:inj_qm}, we have introduced the $c,v$ band resolved $\vec q$-dependent interband quantum geometric tensor as 
\be
Q_{ba}^{cv} (\vec k, \vec q) = r_{cv}^b (\vec k, \vec q) [ r_{cv}^a (\vec k, \vec q) ]^*. 
\ee
For linearly polarised light with polarisation angle $\theta$, the injection current is determined by the real part of $Q_{ba}^{cv} (\vec k, \vec q) $: 
\be
\p_t \vec j^{\rm inj}_\theta (\vec q) = - \frac{e^3\pi }{2 \hbar} \sum_{\vec k, a, b} \rho(\vec k, \vec q) \boldsymbol \Delta (\vec k, \vec q) G_{ba}^{cv} (\vec k, \vec q)  E_b E_a^*
\ee
with $G _{ba}^{cv} (\vec k, \vec q) = \Re [Q_{ba}^{cv} (\vec k, \vec q)]$ the $\vec q$-dependent interband quantum metric. We note parenthentically that the helicity dependent circular injection current is determined by the imaginary part of the $\vec q$-dependent interband quantum geometric tensor multiplied by $\boldsymbol \Delta (\vec k, \vec q)$. At $\vec q=0$, this reduces to interband Berry curvature dipole for vertical transitions; this reproduces the well-known result for quantised circular injection photocurrents~\cite{deJuan2017, Nagaosa2020}.

\section{Symmetry Analysis for Polariton-drag (PD) Shift and Injection Currents}
In this section, we discuss the symmetry properties of photon drag shift and injection currents induced by linearly or circularly polarised light. We will demonstrate that properties of the PD injection and shift photocurrents are sensitive to the symmetry of the material irradiated as well as the light polarisation. These properties can be obtained by examining how the Bloch wavefunction and velocity matrix elements transform under various symmetry operators. 

We begin with the Bloch Hamiltonian $H (\vec k) = e^{-i \vec k \cdot \vec r} \mathcal H(\vec r) e^{i \vec k \cdot \vec r}$. The Bloch wavefunction $|u_n (\vec k) \ra$ in band $n$ satisfies $H(\vec k ) |u_n (\vec k) \ra = \epsilon_{n,\vec k} | u_n (\vec k)\ra$. We proceed by considering how the Bloch hamiltonian and its associated Bloch wavefunctions transform when the material possesses (i) spatial inversion ($\mathcal P$) symmetry [so that $\mathcal P H(\vec k) \mathcal P^{-1} = H(- \vec k)$], or (ii) time-reversal ($\mathcal{T}$) symmetry [so that $\mathcal T H(\vec k) \mathcal T^{-1} = H (- \vec k ) $] respectively.

When the material possesses spatial inversion symmetry, the Bloch hamiltonian obeys
\be
\mathcal P H(\vec k) |u_n (\vec k)\ra = \mathcal P H(\vec k) \mathcal P^{-1} \mathcal P |u_n (\vec k)\ra = \epsilon_{n, \vec k} \mathcal P |u_n (\vec k)\ra = H(-\vec k) \mathcal P |u_n (\vec k)\ra,
\ee 
yielding the following constraints on the energy dispersion and the Bloch wavefunctions:
\be 
\epsilon_{n, \vec k} = \epsilon_{n, - \vec k}, \quad \mathcal P |u_n (\vec k)\ra = C_{n, \vec k} |u_n (- \vec k)\ra,
\ee
where $C_{n, \vec k}$ is a complex phase factor associated with the $\mathcal P$ transformation satisfying $|C_{n, \vec k}| = 1$. Since $\mathcal P$ is unitary and preserves inner product, we have 
\be\label{eq:P_wavefunc}
\la u_{n} (- \vec k_1) | u_m ( -\vec k_2) \ra = \la C_{n, \vec k_1}^* \mathcal P u_{n} ( \vec k_1) | C_{m, \vec k_2}^* \mathcal P u_m ( \vec k_2) \ra = C_{n, \vec k_1} C_{m, \vec k_2}^* \la u_{n} (\vec k_1) | u_m ( \vec k_2) \ra.
\ee

Furthermore, under spatial inversion symmetry, the velocity operator transforms as $\mathcal P \hat{\boldsymbol \nu} \mathcal P^{-1}  = - \hat{\boldsymbol \nu}$. Thus the velocity matrix element satisfies
\begin{align} \label{eq:P_v}
\la u_{n} (  \vec k_1) | \hat{\boldsymbol \nu} | u_m (  \vec k_2) \ra &= \la u_{n} (  \vec k_1) | \mathcal P^{-1} \mathcal P \hat{\boldsymbol \nu} \mathcal P^{-1} \mathcal P | u_m (  \vec k_2) \ra = - C_{n, \vec k_1}^* C_{m, \vec k_2} \la u_{n} (- \vec k_1) |\hat{\boldsymbol \nu} | u_m ( -\vec k_2) \ra.
\end{align}

In the same fashion as above, when the material possesses time-reversal symmetry, the energy dispersion and Bloch wavefunctions transform as
\be 
\epsilon_{n, \vec k} = \epsilon_{n, - \vec k}, \quad \mathcal T |u_n (\vec k)\ra = C_{n, \vec k}' |u_n (- \vec k)\ra,
\ee
where $C_{n, \vec k}'$ is a complex phase factor associated to $\mathcal T$ operation with $|C_{n, \vec k}'|=1$. 
In addition, since $\mathcal T$ is anti-unitary, the inner product of the wavefunctions satisfies 
\be\label{eq:T_wavefunc}
\la u_{n} ( - \vec k_1) | u_m ( - \vec k_2) \ra = \la C_{n, \vec k_1}'^* \mathcal T u_{n} ( \vec k_1) |  C_{m, \vec k_2}'^* \mathcal T u_m ( \vec k_2) \ra = C_{n, \vec k_1}' C_{m, \vec k_2}'^* \la u_{n} ( \vec k_1) | u_m (\vec k_2) \ra^*.
\ee
The velocity operator transforms as $\mathcal T \hat{\boldsymbol \nu} \mathcal T^{-1} = - \hat{\boldsymbol \nu}$, and thus the velocity matrix element satisfies 
\begin{align}\label{eq:T_v}
\la u_{n} ( - \vec k_1) | \hat{\boldsymbol \nu} | u_m ( - \vec k_2) \ra &=  \la C_{n, \vec k_1}'^* \mathcal T u_{n} ( \vec k_1) | \mathcal T (- \hat{\boldsymbol \nu}) \mathcal T^{-1} | C_{m, \vec k_2}'^* \mathcal T u_m ( \vec k_2) \ra =  - C_{n, \vec k_1}' C_{m, \vec k_2}'^* \la u_{n} ( \vec k_1) |  \hat{\boldsymbol \nu} | u_m ( \vec k_2) \ra^* .
\end{align}

The symmetry properties of the Bloch hamiltonian can also be constrained by other point group symmetries of the crystal. A particularly interesting example is that of mirror symmetry. For example, in the presence of mirror symmetry along the $y$-axis, such that $\mathcal M_y H(\vec k) \mathcal M_y^{-1} = H(\mathcal M_y \vec k)$, where $\mathcal M_y: (x,y) \to (-x, y)$. The energy dispersion thus satisfies $\epsilon_{n, \vec k}  = \epsilon_{n, \mathcal M_y \vec k} $. On the other hand, the velocity operator transforms as $\mathcal M_y \hat \nu_x \mathcal M_y^{-1} =  -\hat \nu_x$ and $\mathcal M_y \hat \nu_y \mathcal M_y^{-1} = \hat \nu_y$. Following similar arguments as above, we obtain symmetry relations for wavefunctions and velocity matrix elements in much the same form as above, leading to distinctive properties of the PD injection and shift photocurrents as discussed in the main text and below. 

\subsection{Symmetry analysis for PD injection current}

\subsubsection{Inversion symmetry} 

When the material possesses $\mathcal{P}$-symmetry and identifying band indices $m, \; n$ in Eq.~(\ref{eq:P_v}) with $c , \; v$, we find that the interband velocity matrix element $\boldsymbol \nu_{cv} (\vec k, \vec q)$ satisfies $ \boldsymbol \nu_{cv} (\vec k, \vec q) = - C_{c, \vec k+ \vec q/2}^* C_{v, \vec k - \vec q/2} \boldsymbol \nu_{cv} (-\vec k, - \vec q)$. Thus, for linear [denoted as $\theta$] and circularly [denoted as $\eta = \pm 1$] polarised light, the square of the transition matrix element
 \be 
 \mathfrak{v}_{cv}^{\theta (\eta)} (\vec k, \vec q) = |\vec E^{\theta (\eta)} \cdot \boldsymbol \nu_{cv} (\vec k, \vec q)|^2
 \ee 
 thus obeys $\mathfrak v_{cv}^{\theta (\eta)} (\vec k, \vec q) = \mathfrak v_{cv}^{\theta (\eta)} (- \vec k, -\vec q)$.

Next we note when the material possesses $\mathcal{P}$-symmetry, the group velocities in valence and conduction bands satisfy $\vec v_{c} (\vec k+ \vec q/2) = -\vec v_{c} (- \vec k - \vec q/2)$ and $\vec v_{v} (\vec k- \vec q/2) = -\vec v_{v} (- \vec k + \vec q/2)$, we have $\boldsymbol \Delta (\vec k, \vec q) = -\boldsymbol \Delta (- \vec k, -\vec q)$ odd under $\vec k \to - \vec k, \; \vec q \to - \vec q$. On the other hand, since $\epsilon_{n, \vec k}$ is even in k-space, we have $\rho (\vec k, \vec q) = \rho (-\vec k, -\vec q)$. 

Therefore, in the presence of inversion symmetry, the injection current (obtained by summing Eq.~(\ref{eq:inj}) of the main text over $k$-space) obeys 
\be \label{eq:P_inj}
\p_t \vec j^{\rm inj}_\theta (\vec q) = - \p_t \vec j^{\rm inj}_\theta (-\vec q), \quad \p_t \vec j^{\rm inj}_{\rm cir} (\vec q) = - \p_t \vec j^{\rm inj}_{\rm cir} (-\vec q),
\ee
as discussed in Table I of the main text.

\subsubsection{Time reversal symmetry}

When the material possesses $\mathcal{T}$-symmetry, Eq.~(\ref{eq:T_v}) gives $\boldsymbol \nu_{cv} (\vec k ,\vec q) = - C_{c, \vec k + \vec q/2}' C_{v, \vec k - \vec q/2}'^* \left[ \boldsymbol \nu_{cv} (- \vec k, - \vec q) \right]^*$. For linearly polarised light, since $\vec E^\theta = (\vec E^\theta)^*$, we have $\mathfrak v_{cv}^{\theta} (\vec k, \vec q) = \mathfrak v_{cv}^{\theta} (- \vec k, -\vec q)$. In contrast, for circularly polarised light, we have $\mathfrak v_{cv}^{\eta} (\vec k, \vec q) = \mathfrak v_{cv}^{-\eta} (- \vec k, -\vec q)$. This latter relation can be obtained by noting $\vec E^{\eta} = (\vec E^{-\eta})^*$ for circularly polarised irradiation. 

We now turn to the carrier velocity $\vec v_{c(v)} (\vec k)$. For $\mathcal{T}$-symmetry preserving materials, we have $\vec v_{c} (\vec k+ \vec q/2) = -\vec v_{c} (- \vec k - \vec q/2)$ and $\vec v_{v} (\vec k- \vec q/2) = -\vec v_{v} (- \vec k + \vec q/2)$. Thus, the change in carrier velocity obeys $\boldsymbol \Delta (\vec k, \vec q) = -\boldsymbol \Delta (-\vec k, -\vec q)$. Similar to that discussed above for inversion symmetry, $\mathcal{T}$-symmetry preserving materials also possess energy dispersion relations that are even in $k$-space yielding $\rho (\vec k, \vec q) = \rho (-\vec k, -\vec q)$. 

As a result, the linear and circular injection current (obtained by summing Eq.~(\ref{eq:inj}) of the main text over $k$-space) obeys
\be
\p_t \vec j^{\rm inj}_\theta (\vec q) = - \p_t \vec j^{\rm inj}_\theta (-\vec q), \quad \p_t \vec j^{\rm inj}_{\rm cir} (\vec q) =  \p_t \vec j^{\rm inj}_{\rm cir}  (-\vec q)
\label{eq:T_inj}
\ee
as discussed in Table 1 of the main text. 

Combining both Eq.~(\ref{eq:P_inj}) and Eq.~(\ref{eq:T_inj}), we conclude that PD linear injection charge photocurrents are in general allowed in materials with both $\mathcal P$- and $\mathcal T$-symmetries. In contrast, PD circular injection charge photocurrents (a photocurrent that depends on the helicity of the incident light) vanishes when both $\mathcal P$- and $\mathcal T$-symmetries in the material remain intact.

\subsubsection{Mirror symmetry}

It is also interesting to consider how point group symmetries can also similarly constrain the form of the PD injection photocurrents. As a simple illustration we focus on PD linear injection photocurrents in a material with a mirror axis along $y$. For simplicity, we consider the case where incident light (linear) polarization [$\vec E = E_0 \hat{\vec y}$] as well as non-vertical transition wavevector $\vec q$ is directed along the mirror axis $y$. As a result, for $\vec q = q \hat{\vec y}$, we have $\mathcal M_y \vec q= \vec q$. In this case, the square of the transition matrix element obeys $\mathfrak v_{cv}^{\theta}  (\vec k, \vec q) = \mathfrak v_{cv}^{\theta} (\mathcal M_y \vec k, \vec q)$. On the other hand, the component of the change in electron group velocity normal to the mirror axis will switch sign $\Delta_x (\vec k, \vec q) = -\Delta_x (\mathcal M_y \vec k, \vec q)$ while the component parallel to the mirror axis remains invariant $\Delta_y (\vec k, \vec q) = \Delta_y (\mathcal M_y \vec k, \vec q)$ under mirror reflection. 

\begin{figure} 
    \centering
    \includegraphics[scale=0.6]
{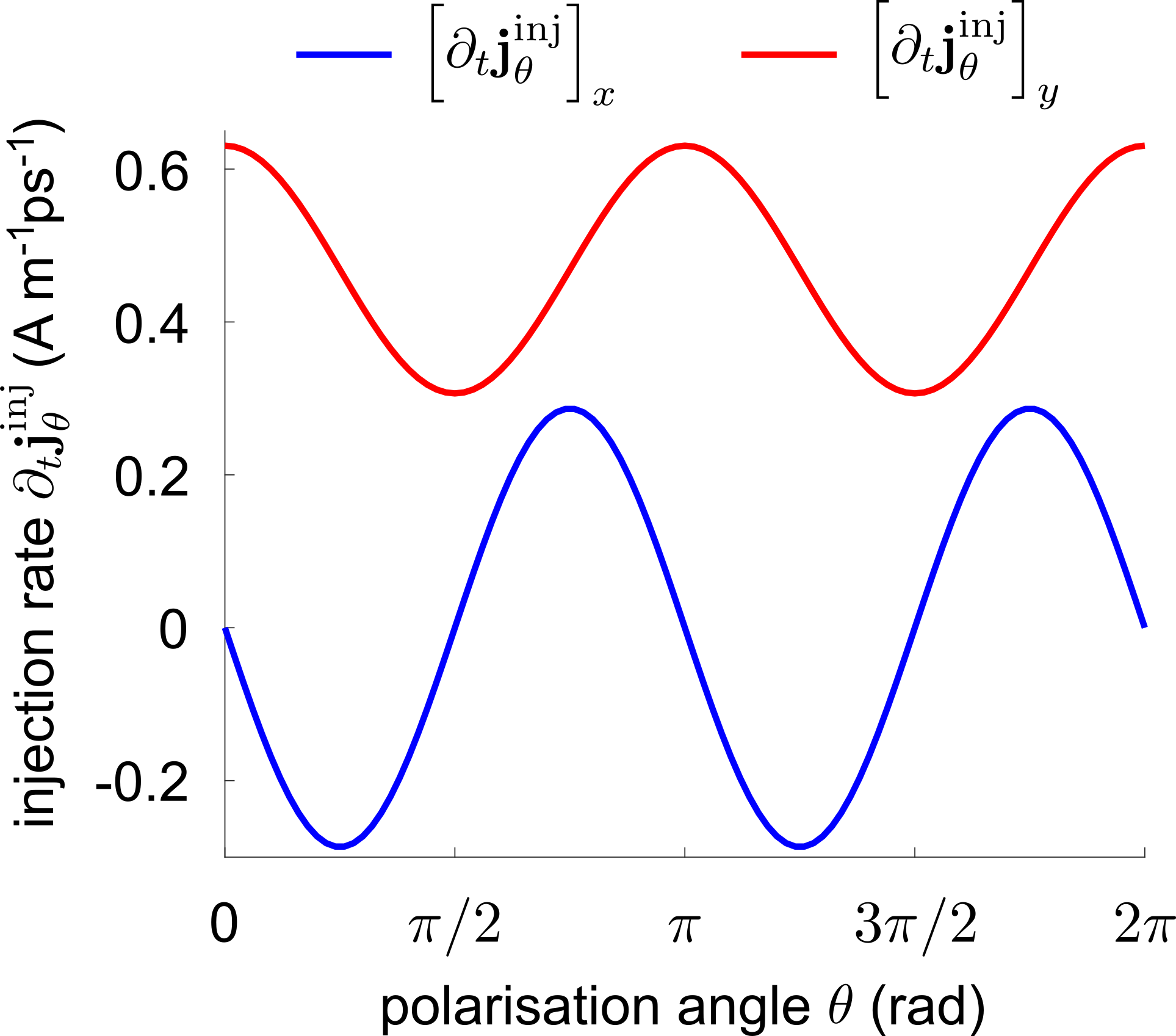}
    \caption{Polariton drag linear injection photocurrent as a function of light polarisation angle in centrosymmetric BLG. The plasmon polariton wavevector is fixed at $|\vec q | = 0.03 \; {\rm nm^{-1}}$, and all other parameters are the same as Fig.~\ref{fig2} in the main text. } 
    \label{figS0}
\end{figure}

We note that $\rho (\vec k, \vec q) = \rho (\mathcal M_y \vec k, \vec q) $ only depends on the energy dispersion (which is even under mirror reflection). As a result, the component of the PD linear injection photocurrent normal to $\vec q$ (when it is directed along the mirror axis) vanishes: $[\p_t \vec j^{\rm inj}_\theta]_x (\vec q) = 0$, while the component parallel to $\vec q$ when it is directed along the mirror axis, $[\p_t \vec j^{\rm inj}_\theta]_y (\vec q)$, is allowed. This is verified in Fig.~\ref{figS0}, which plots the linear injection photocurrent as a function of light polarisation angle $\theta$ for a fixed $\vec q $ along the $y$-direction (i.e the mirror axis). We observe that the linear injection current flows along the mirror plane when the electric field is polarised along or perpendicular to the mirror plane.

\subsection{Symmetry analysis for PD shift photocurrent}

\subsubsection{Inversion symmetry}

When the material possesses $\mathcal{P}$-symmetry, the inner product of the wavefunctions follows the relation in Eq.~(\ref{eq:P_wavefunc}) while the velocity matrix element obeys Eq.~(\ref{eq:P_v}). Since all the Bloch wavefunctions in Eq.~(\ref{eq:Wilson}) occur in pairs (guaranteeing its gauge invariance), the phase factors for the wavefunctions resulting from the $\mathcal P$ transformation fully compensate with each other. As a result, we find 
\begin{align}
&\mathcal W (\vec k, \delta \vec k, \vec q) = - \mathcal W (-\vec k, -\delta \vec k, -\vec q), \quad  \arg \left[ \mathcal W (\vec k, \delta \vec k, \vec q)  \right] = \arg \left[ \mathcal W (-\vec k, -\delta \vec k, -\vec q) \right] + \pi.
\end{align}
Since the shift vector $\vec r (\vec k, \vec q)$ depends on the derivative of $\arg \left[ \mathcal W(\vec k, \delta \vec k, \vec q) \right]$, we arrive at 
\be
\vec R^{\theta (\eta)} (\vec k, \vec q) = - \vec R^{\theta(\eta)} (- \vec k , - \vec q),
\ee
where the shift vector flips direction under $\vec k \to - \vec k, \; \vec q \to - \vec q$. 

To understand the symmetry properties of the PD shift charge photocurrent, we recall that both $\mathfrak v_{cv}^{\theta (\eta)} (\vec k, \vec q)$ and $\rho (\vec k, \vec q)$ are even under $\vec k \to - \vec k, \; \vec q \to - \vec q$ [see above]. By summing Eq.~(\ref{eq:shift}) of the main text over $k$-space, we find that the PD shift photocurrents flow in opposite directions for $\pm \vec q$ in $\mathcal{P}$-preserving materials:
\be\label{eq:P_shift}
\vec j^{\rm shift}_\theta (\vec q) = - \vec j^{\rm shift}_\theta (-\vec q), \quad \vec j^{\rm shift}_{\rm cir} (\vec q) = - \vec j^{\rm shift}_{\rm cir} (-\vec q).
\ee 
as shown in Table I of the main text.

\subsubsection{Time reversal symmetry}

Following similar analysis for the injection current, in $\mathcal T$-symmetry preserving materials, we have $\hat{\vec e}_\theta  \cdot  \boldsymbol \nu_{cv} (\vec k, \vec q) = - C_{c,  \vec k + \vec q/2}' C_{v, \vec k - \vec q/2}'^* \left[\hat{\vec e}_\theta  \cdot \boldsymbol \nu_{cv} (- \vec k,  - \vec q) \right]^*$ for linearly polarised light. Combining with the relation for the Bloch wavecfunctions in Eq.~(\ref{eq:T_wavefunc}), we obtain 
\begin{align}
&\mathcal W^\theta (\vec k, \delta \vec k, \vec q) = -\left[ W^\theta (-\vec k, -\delta \vec k, -\vec q) \right]^*,  \quad \arg \left[ \mathcal W^{\theta} (\vec k, \delta \vec k, \vec q) \right] = - \arg \left[ \mathcal W^{\theta} (-\vec k, -\delta \vec k, -\vec q) \right] + \pi.
\end{align}
Here we note that the additional phase factors $C'$ that arise under $\mathcal T$ transformation fully compensate each other since the Bloch wavefunctions in Eq.~(\ref{eq:Wilson}) occur in pairs. 

By taking the derivative of the phase of $\mathcal W^\theta(\vec k, \delta \vec k, \vec q)$, we arrive at the symmetry constraint for the shift vector (for linearly polarized light) in $\mathcal{T}$-preserving materials: 
\be
\vec R^{\theta} (\vec k, \vec q) =  \vec R^{\theta} (- \vec k , - \vec q), 
\ee
where the shift vector is even under $\vec k \to - \vec k, \; \vec q \to - \vec q$.

On the other hand, for circularly polarised light, we have $\hat{\vec e}^\eta  \cdot \boldsymbol \nu_{cv} (\vec k, \vec q) = -C_{c,  \vec k+ \vec q/2}' C_{v, \vec k - \vec q/2}'^*  \left[\hat{\vec e}^{-\eta}  \cdot \boldsymbol \nu_{cv} (- \vec k,  - \vec q) \right]^*$. Similarly, the Wilson loop satisfies 
\begin{align}
\mathcal W^{\eta} (\vec k, \delta \vec k, \vec q) = - \left[ \mathcal W^{-\eta} (-\vec k, -\delta \vec k, -\vec q) \right]^* , \quad \arg \left[ \mathcal W^{\eta} (\vec k, \delta \vec k, \vec q) \right] = - \arg \left[ \mathcal W^{-\eta} (-\vec k, -\delta \vec k, -\vec q) \right] + \pi.
\end{align}

As a result, we find that the shift vector (for circularly polarized light with helicity $\eta$) satisfies 
\be
\vec R^{\eta} (\vec k, \vec q) =  \vec R^{-\eta} (- \vec k , - \vec q). 
\ee

By summing {Eq.~(\ref{eq:shift}) of the main text over $k$-space,} the PD linear and circular shift charge photocurrents in $\mathcal T$-symmetric materials obey 
\be\label{eq:T_shift}
\vec j^{\rm shift}_\theta (\vec q) = \vec j^{\rm shift}_\theta ( - \vec q), \quad \vec j^{\rm shift }_{\rm cir} (\vec q) =  - \vec j^{\rm shift }_{\rm cir} ( - \vec q) .
\ee

Combining with the constraints in Eq.~(\ref{eq:P_shift}) and~(\ref{eq:T_shift}), we find that in $\mathcal P$- and $\mathcal T$-symmetric materials, PD linear shift charge photocurrent vanishes for all non-vertical wavevectors $\vec q$ while PD circular shift charge photocurrent are allowed.

\begin{figure} 
    \centering
    \includegraphics[scale=0.53]
{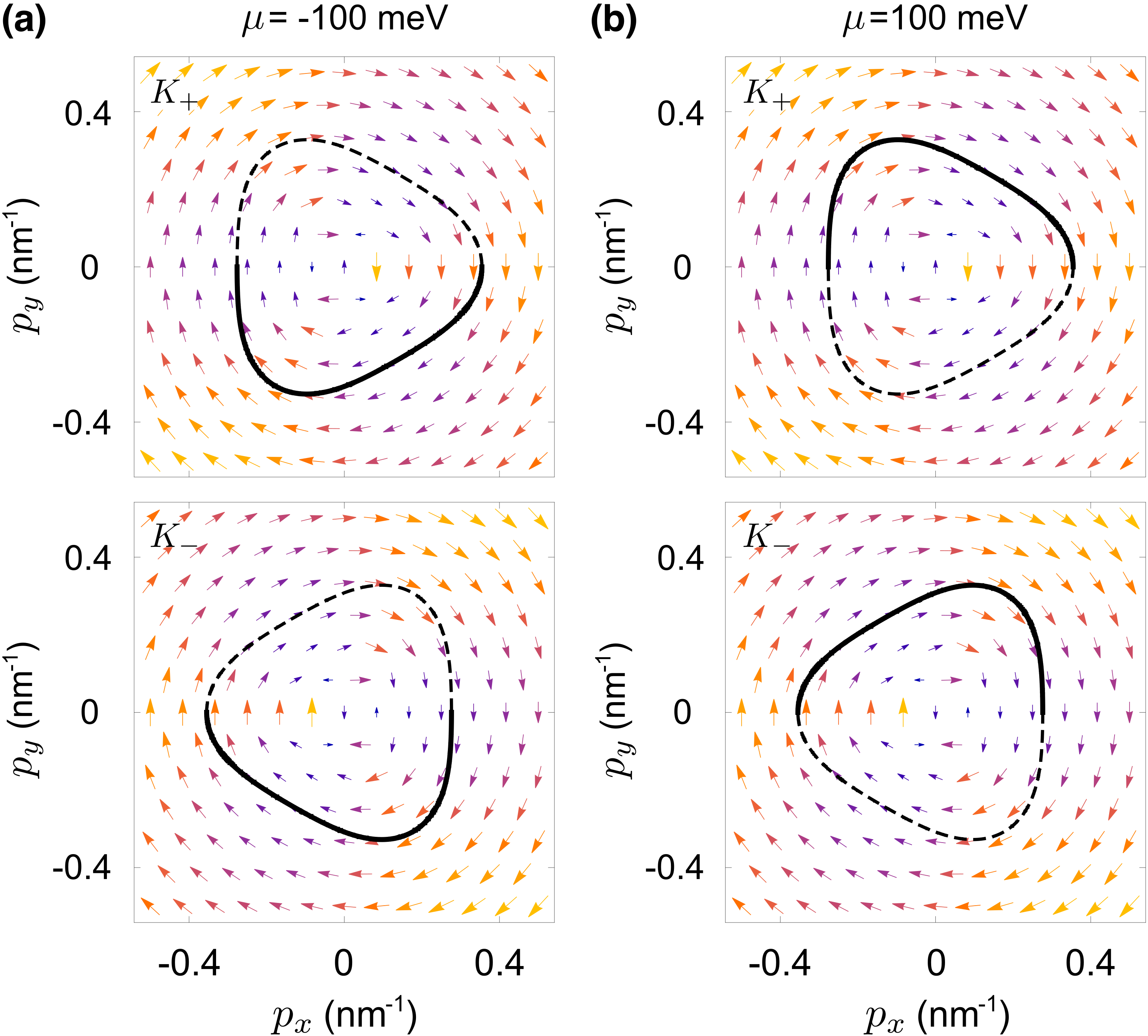}
    \caption{Polariton selective photoexcitation (PSP) of charge carriers near the Fermi surface for non-vertical interband transitions and circular shift vector. Importantly, PSP yields an imbalanced sampling of shift vector when carriers close to the Fermi surface are excited. Weighted PD circular shift vector $\tilde{\vec R}^{\zeta, \eta} (\vec p, \vec q)$ for $\eta =+1$ with contour line plots (black) indicating the regions that satisfy energy and momentum conservation. Solid line indicate shift vector regions that are sampled, dashed indicate regions that are not sampled when chemical potential is fixed at $ \mu = - \hbar \omega/2$ (a) and $\mu = \hbar \omega/2$ (b) in the $K_+$ (top panel) and $K_-$ (bottom panel) valleys. Parameters used are the same as Fig.~\ref{fig1} of the main text.} 
    \label{figS1}
\end{figure}

\subsubsection{Mirror symmetry} 

Here we illustrate how mirror symmetry can constrain the form of the PD circular shift photocurrent. As a simple illustration we focus on a mirror plane axis along $y$  and consider non-vertical transition wavevector $\vec q$ directed along the mirror axis $y$.

For circularly polarised light with polarisation vector $\hat{\vec e}^\eta$, the square of the transition matrix element satisfies $\mathfrak v_{cv}^\eta (\vec k, \vec q) = \mathfrak v_{cv}^{-\eta} (\mathcal M_y \vec k, \vec q) $. The Wilson loop obeys 
\be
\mathcal W^{\eta} (\vec k, \delta \vec k, \vec q) =  - \mathcal W^{-\eta} (\mathcal M_y \vec k, \mathcal M_y \delta \vec k, \vec q)  , \quad \arg \left[ \mathcal W^{\eta} (\vec k, \delta \vec k, \vec q)  \right] =  \arg \left[ \mathcal W^{-\eta} (\mathcal M_y \vec k, \mathcal M_y \delta \vec k, \vec q) \right]  + \pi .
\ee
By taking the derivative with respect to $\delta \vec k$, we have
\be
R_x^{\eta} (\vec k, \vec q) = - R_x^{-\eta} (\mathcal M_y \vec k, \vec q), \quad R_y^{\eta} (\vec k, \vec q) =  R_y^{- \eta} (\mathcal M_y \vec k, \vec q). 
\ee

Since $\rho (\vec k, \vec q ) = \rho( \mathcal M_y \vec k, \vec q)$, the $y$-component of the helicity dependent charge circular shift current vanishes while $\left[ \vec j^{\rm shift}_{\rm cir} \right]_x$ is allowed, i.e. PD charge circular shift current in the presence of $\mathcal M_y$-symmetry is purely transverse.

\section{Giant enhancement of circular shift photocurrent due to PSP}

Non-vertical transitions enable polariton selective photoexcitation of charge carriers near the Fermi surface. In particular, when $\mu = \pm \hbar \omega /2$, only half of the interband transition contour (defined by $\delta (\epsilon_{cv}(\vec k, \vec q) - \hbar \omega)$ can be photoexcited, leading to giant enhancement in photocurrents. As shown by Fig.~\ref{fig1} of the main text, the PD circular shift photocurrents exhibits large and opposite peaks at $\mu = \pm \hbar \omega$. To visualise this PSP induced resonance effect, we plot the distribution of the weighted shift vector $\tilde{\vec R}^{\zeta,\eta} (\vec p,\vec q) = \mathfrak v^{\zeta,\eta} (\vec p, \vec q) \vec R^{\zeta, \eta} (\vec p, \vec q)$ [this determines the direction of the PD CS photocurrent, see Eq.~(\ref{eq:shift})] in Fig.~\ref{figS1}. Here the interband transition contours (black) indicate $\vec p$ values that satisfy $\delta (\epsilon_{cv}^\zeta (\vec p, \vec q) - \hbar \omega)$. When $\mu = - \hbar \omega/2$ (in the valence band), the Fermi surface intersects with the interband transition contour so that only the bottom half of the transition contour contributes to the non-vertical interband transitions (solid curve in Fig.~\ref{figS1}a). These $\vec p$ values correspond to occupied carriers in the valence band so that $f_{cv} (\vec p, \vec q) \neq 0$. In contrast, the other half (dashed curve) do not contribute to the non-vertical interband transitions ($f_{cv} (\vec p, \vec q) = 0$). This asymmetric sampling of charge carriers on the interband transition contour (enforced by the occupation factors) avoids cancellation of $\tilde{\vec R}^{\zeta, \eta}(\vec p , \vec q)$ in k-space, leading to large PD CS photocurrents.   Similarly, when $\mu = \hbar \omega/2$ is in the conduction band (Fig.~\ref{figS1}b), only the top half of the transition contour is available for interband transitions (solid curve in Fig.~\ref{figS1}b); these $\vec p$ values correspond to the region of the conduction band that is unoccupied thus allowing interband transitions ($f_{cv} (\vec p, \vec q) \neq 0$). Comparing the directions of the weighted shift vector, this yields an PSP enhanced $\vec j_{\rm cir}^{\rm shift}$ that switches sign when the Fermi energy is moved from valence band to the conduction band.

\section{$c,v$ band symmetry and PD photocurrents in bilayer graphene}

In this section, we discuss how a symmetry between the conduction $c$ and valence $v$ bands can emerge in the low-energy effective hamiltonian for bilayer graphene. As we will show below, this effective $c,v$ band symmetry leads to a vanishing PD linear injection and circular shift photocurrents at low temperature when the Fermi surface (determined by $\mu$) does not intersect and are far from the interband transition contours (determined by $\hbar \omega$). 

We consider the low energy Hamiltonian in Eq. (\ref{eq:H}) in the main text. For a two-band Hamiltonian, we can directly solve for the eigenenergies and eigenstates. As discussed above, these enable to directly compute the shift vector $\vec r^{\zeta,\eta} (\vec p, \vec q)$ and the change in carrier velocity $\boldsymbol \Delta^\zeta (\vec p, \vec q)$ as a carrier is photoexcited between $c$ and $v$ bands. For the convenience of the reader, we rewrite $H(\vec p)$ as 
\be
H(\vec p) = d_1 \sigma_x + d_2 \sigma_y, \quad d_1 = - \frac{\hbar^2}{2 m} (p_x^2 - p_y^2) + \hbar v_3 \zeta p_x , \quad d_2 = - \frac{\hbar^2}{ m} \zeta p_x p_y - \hbar v_3 p_y.
\label{eq:section3H}
\ee
The energy dispersion is given by $\epsilon_{c,v}^\zeta (\vec p) = \pm \sqrt{d_1^2 + d_2^2}$, where the explicit $\zeta$- and $\vec p$- dependence of $d_1$ and $d_2$ is suppressed for brevity. The corresponding eigenstates are 
\be \label{eq:wavefuc}
| u_c^\zeta (\vec p) \ra = \frac{1}{\sqrt{2}} \begin{pmatrix}
e^{- i \phi_{\zeta, \vec p} } \\
1 \end{pmatrix}, \quad | u_v^\zeta (\vec p) \ra =   \frac{1}{\sqrt{2}} \begin{pmatrix}
e^{- i \phi_{\zeta, \vec p} } \\
-1  \end{pmatrix},
\ee
where $\phi_{\zeta, \vec p} = \tan^{-1} \left( d_2 / d_1 \right)$. As we now show, for $\mathcal P$- and $\mathcal T$-symmetric bilayer graphene, there is an additional emergent symmetry between the conduction and the valence bands that relates the conduction and valence bands in the separate valleys. 

To see this, first we note that (as can be verified by inspection) the energies of the conduction and valence bands in the separate valleys obey $\epsilon_c^\zeta (\vec p) = - \epsilon_v^\zeta (\vec p) = - \epsilon_v^{-\zeta} (-\vec p)$. This emergent $c,v$ band symmetry yields a non-vertical interband transition energy: $\epsilon_{cv}^\zeta (\vec p, \vec q)  = \epsilon_c^\zeta (\vec p+\vec q/2) - \epsilon_v^\zeta (\vec p-\vec q/2)$ that obeys
 \be
 \epsilon_{cv}^\zeta (\vec p, \vec q) = \epsilon_{cv}^{-\zeta} (- \vec p, \vec q)
 \ee

This symmetry between $c$ and $v$ bands between valleys also constrains the interband velocity matrix elements. Using Eq.~(\ref{eq:wavefuc}), we can explicitly compute $\boldsymbol \nu_{cv}^\zeta (\vec p, \vec q) = \la u_c^\zeta (\vec p+ \vec q/2) | \hat{\boldsymbol \nu}  | u_v^\zeta (\vec p - \vec q/2) \ra$ [where $\hat{\boldsymbol \nu} = \nabla_{\vec p} H(\vec p) /\hbar$] as 
\begin{align}
\nu_{cv,x}^\zeta (\vec p, \vec q) &= \frac{1}{2} \left( - \frac{\hbar}{m} p_x + v_3 \zeta \right)  \left[ - e^{i \phi_{\zeta, \vec p+ \vec q/2}} + e^{-i \phi_{\zeta, \vec p - \vec q/2}}  \right] - \frac{i}{2} \frac{\hbar}{m} \zeta p_y  \left[ e^{i \phi_{\zeta, \vec p+ \vec q/2}} + e^{-i \phi_{\zeta, \vec p - \vec q/2}} \right] \nonumber \\
\nu_{cv,y}^\zeta (\vec p, \vec q) &= \frac{1}{2} \frac{\hbar}{m} p_y  \left[ - e^{i \phi_{\zeta, \vec p+ \vec q/2}} + e^{-i \phi_{\zeta, \vec p - \vec q/2}}  \right] - \frac{i}{2} \left( \frac{\hbar}{m} \zeta p_x  + v_3 \right)  \left[ e^{i \phi_{\zeta, \vec p+ \vec q/2}} + e^{-i \phi_{\zeta, \vec p - \vec q/2}} \right]
\end{align}
We note that when $\zeta \to - \zeta, \; \vec p\to - \vec p$, we have $d_1  \to d_1$ and $d_2 \to - d_2$ [see Eq.~(\ref{eq:section3H})]. Thus, $\phi_{\zeta, \vec p} = - \phi_{-\zeta, -\vec p}$ is odd in k-space, and we have
 \be
 \boldsymbol \nu_{cv}^\zeta (\vec p, \vec q) = \boldsymbol \nu_{cv}^{-\zeta} (-\vec p, \vec q). 
 \label{eq:section3v}
 \ee
 As a result, the square of the interband transition matrix for linearly (circularly) polarised light obeys $\mathfrak v^{\zeta, \theta (\eta)} (\vec p, \vec q) = \mathfrak v^{-\zeta, \theta (\eta)} (-\vec p, \vec q)$.

The above symmetry relations for how velocity matrix element (and the energies) transform as $\zeta \to - \zeta, \; \vec p\to - \vec p$ can be directly used to determine the the PD circular shift photocurrent. To proceed, we consider the shift vector for circularly polarized light reproduced here for the convenience of the reader as 
\be
\vec R^{\zeta,\eta} (\vec p, \vec q) = [\vec A_c^\zeta (\vec p + \vec q/2) - \vec A_v^{\zeta} (\vec p - \vec q/2)] - \nabla_{\vec p} \arg \left( \vec E^\eta \cdot \boldsymbol \nu^{\zeta}_{cv} (\vec p, \vec q) \right)
\label{eq:section3r}
\ee 
By direct computation using Eq.~(\ref{eq:wavefuc}), we find the Berry connection in the valence and conduction bands in opposite valleys satisfy $\vec A_c^\zeta (\vec p) = \vec A_v^{-\zeta} (- \vec p)$. This means that the difference of Berry connections [square brackets in Eq.~(\ref{eq:section3r})] is odd when $\zeta \to - \zeta, \; \vec p\to - \vec p$, namely: $\vec A_c^\zeta (\vec p + \vec q/2) - \vec A_v^{\zeta} (\vec p - \vec q/2) = - \left[ \vec A_c^{-\zeta} ( - \vec p + \vec q/2) - \vec A_v^{-\zeta} (-\vec p - \vec q/2) \right]$. Further, by applying Eq.~(\ref{eq:section3v}) to the last term of Eq.~(\ref{eq:section3r}) we find: $\nabla_{\vec p} \arg \left( \vec E^\eta \cdot \boldsymbol \nu^{\zeta}_{cv} (\vec p, \vec q) \right)$ is also odd as $\zeta \to - \zeta, \; \vec p\to - \vec p$. 
Hence, due to the emergent $c,v$ band symmetry, we find that as $\zeta \to - \zeta, \; \vec p\to - \vec p$ the weighted shift vector $\tilde{\vec R}^{\zeta, \eta} (\vec p, \vec q) \equiv \mathfrak v^{\zeta, \theta (\eta)} (\vec p, \vec q) \vec R^{\zeta, \eta} (\vec p, \vec q)$ obeys 
\be
\tilde{\vec R}^{\zeta, \eta} (\vec p, \vec q) = - \tilde{\vec R}^{-\zeta, \eta} (-\vec p, \vec q)
\ee
This is verified in Fig.~\ref{figS1}, which shows the numerical vector plot for $\tilde{\vec R}^{\zeta, \eta} (\vec p, \vec q)$. Finally, we note that when the Fermi surface is far from any interband transition contours such that $f_{cv}^\zeta (\vec p, \vec q)$ is a constant for all $\vec p$, we have
\be
\rho^\zeta (\vec p, \vec q) = \rho^{-\zeta} (-\vec p, \vec q)
\label{eq:section3rho}
\ee 
This can be achieved, for instance, for a large $\hbar \omega$ and chemical potential fixed close to charge neutrality at low temperature. In this case, by summing the expression for the PD circular shift photocurrent in Eq.~(\ref{eq:shift}) of the main text across all $\vec p$ and both valleys, we find 
the PD circular shift photocurrent vanishes due to the emergent symmetry between the valence and the conduction bands. 

A similar argument can also be applied to the PD linear injection photocurrent. The change in electron group velocity $\boldsymbol \Delta^\zeta(\vec p, \vec q)$ can be written as $\boldsymbol \Delta^\zeta(\vec p, \vec q) = \nabla_{\vec p} \epsilon_{cv}^\zeta (\vec p, \vec q) /\hbar$. Since $\epsilon_{cv}^\zeta (\vec p, \vec q) = \epsilon_{cv}^{-\zeta} (-\vec p, \vec q)$, we have $\boldsymbol \Delta^\zeta(\vec p, \vec q) = - \boldsymbol \Delta^{-\zeta} (- \vec p, \vec q) $. In the same fashion as discussed above, when the Fermi surface is far from any interband transition contours such that $f_{cv}^\zeta (\vec p, \vec q)$ is a constant for all $\vec p$, we have Eq.~(\ref{eq:section3rho}). As a result, in such a situation, applying Eq.~(\ref{eq:section3v}), (\ref{eq:section3rho}), as well as $\boldsymbol \Delta^\zeta(\vec p, \vec q) = - \boldsymbol \Delta^{-\zeta} (- \vec p, \vec q)$, and summing the expression for the PD linear injection photocurrent in Eq.~(\ref{eq:inj}) of the main text across all $\vec p$ and both valleys, we find a vanishing PD linear injection current. 

This emergent $c,v$ band symmetry can be broken in two ways. As we illustrate in the main text, placing the Fermi energy in the valence band or conduction band in the vicinity of $\pm \hbar \omega/2$ naturally breaks the symmetry between the conduction and the valence band, leading to $f_{cv}^\zeta (\vec p, \vec q)$ that is only nonzero for a selective region in the momentum space. This is the polariton selective photoexcitation (PSP) case discussed in the main text. Another way to break the $c,v$ band symmetry is to include a particle-hole asymmetric term in the Bloch Hamiltonian itself, for example, by considering the next-nearest-neighbour hopping in monolayer graphene~\cite{Maysonnave}.

\end{document}